# THE PHYSICS OF FLAVOR IS THE FLAVOR OF PHYSICS[a]

H. FRITZSCH

*Ludwig–Maximilians–Universität München, Sektion Physik, Theresienstraße 37,*
*D–80333 München, Germany*
*and*
*CERN, CH-1211 Geneva 23*

Summary Talk: International Conference of Flavor Physics (ICFP 2001). Zhang–Jia–Jie, Hunan, China (May / June 2001)

For me it is a pleasure to give the Summary Talk at this conference on flavor physics, held near one of the most spectacular places on our planet, the natural wonders of Wulingyan in the Hunan province of China. In the Natural Park we have seen a very spectacular abundance of form and structure. Wulingyuan proves once again that nature prefers complexity instead of simplicity, once it is given the choice. The German philosopher Leibniz suggested more than 300 years ago that we live in the best of all possible worlds. I doubt whether this is true, but certainly he did not mean that we live in the simplest of all possible worlds. It is not the world of Konfuzius, following rigid laws and simple structures, but rather the world of Laotse, full of complexity, a world in the eternal swing between Yin and Yang.

The topic of this conference, the physics of flavor, fits very well into these surroundings. In all of particle physics, the physics of flavor sticks out as the field which has the highest complexity and the richest phenomenology. Indeed, the flavor of particle physics can be seen most clearly by looking at the physics of flavor, with all its phenomena ranging from the spectra of heavy mesons and baryons, from particle–antiparticle oscillations, from $CP$ violation etc. up to exotic pheonomena like neutrino oscillations.

Flavor physics is an area which has emerged as an independent field of high energy physics only after the Standard Model of today had come up in its first contures, shortly after the beginning of the 70'ies. At that time it became apparent that quarks carry color, and the color force is of crucial importance for the understanding of the strong interaction phenomena. Thus the need to distinguish between the color index and the index describing the various types of quarks $u, d, s \dots$ suddenly was in the air, at least at CALTECH, which was about the only place where such subtleties were discussed at that time. For my own private use, I denoted the various quarks as quark types, a notation, I still use today in the German language. Once Gell–Mann and I were driving to a lunch restaurant in Pasadena and passed by a Baskin and Robins icecream place, advertising 32 different flavors. Murray suddenly came up with the proposal to use the name "flavor". I did not like this proposal at the beginning, translating it into German, where it means "Geschmack", an expression one could hardly use for the description of a subatomic particle. Soon afterwards, however, I went along with it, especially after realizing that in other languages the translation of "flavor" gives quite meaningful results. For example, in Italian the word "il sapore", used e. g. in "il sapore del vino" could very well be used to distinguish the various degrees of freedom of the quarks.

---
[a]Supported in part by VW–Stiftung Hannover (I–77495)

Compared to the present time, the flavor physics in those days was rather poor. Only three flavors, i. e. $u, d, s$, were known, and the basic parameters of flavor physics were the three quark masses and the Cabibbo angle. $CP$–violation was considered to be a peculiar phenomenon not intrinsically related to the flavor mixing.

Today we see the sharp contures of the Standard Model[1] in front of us, like the contures of the Wulingyan mountains seen from the Golden Whip Stream. The physics of flavor is at the same time the physics of the multitude of the free parameters of the theory. Even if we disregard possible neutrino masses, the minimal number of parameters is 18, among them the six masses of the quarks, the three lepton masses, and four flavor mixing parameters. Especially those 13 parameters are in the focus of flavor physics. In the Standard Model they arise in a way, which can hardly be considered satisfactory, even on low standards. They appear as the result of a direct coupling of the fermions to the "Higgs" field, a formal device without any predictive power, as far as those parameters are concerned.

In my view, this mechanism of fermion mass generation is the least attractive corner of the Standard Model, and it is quite likely that this is the corner where the model might deviate from reality. Furthermore it might well be that the "Higgs" particle responsible for the generation of mass for the $W$– and the $Z$–bosons does not couple to the $b$–quark with a strength proportional to $m_b$ as expected in the Standard Model, in which case the "Higgs" particle would not decay predominantly into a $\bar{b}b$–system, but into other particles, e. g. into two gluons or into $\gamma\gamma$ (see e. g. ref. (2)).

More than a year ago we have entered the new millenium with a rather bizarre spectrum of the lepton and quark masses, which extends (in the absence of neutrino masses) from about 0.5 MeV (electron mass) to about 175000 MeV ($t$–mass), stretching over almost six orders of magnitude. On a logarithmic scale, the quark masses are nearly on straight lines, if plotted as functions of the family index, implying that the mass ratios are identical:

$$m_c : m_t = m_u : m_c$$
$$m_s : m_b = m_d : m_s \tag{1}$$

The hierarchy exhibited by the mass spectrum is impressive. Moreover, about 97% of the mass is provided by the $t$–mass. The $t$–quark is the only fermion whose mass is comparable to the mass scale of the electroweak symmetry breaking, parametrized by the v.e.v. of the "Higgs" field

$$v \cong 246 \ \ \text{GeV}. \tag{2}$$

The observed $t$–mass is very close to

$$v/\sqrt{2} \cong 174 \ \ \text{GeV} \ \ i.e. \ \ v/m_t \cong \sqrt{2} \tag{3}$$

Thus far this factor $\sqrt{2}$, which looks like a Clebsch–Gordon coefficient, has not been understood; it might, of course, simply be an accident.

Nevertheless the lepton–quark mass spectrum exhibits simple features which ask for a deeper understanding, beyond the rather shallow interpretation given within the Standard Model. Slightly more than 100 years ago the energy spectra observed e. g. for the hydrogen atom found their theoretical explanation within quantum theory. One can expect that in a similar way the fermion mass spectrum is a clear sign that there is physics activity beyond the frontier line drawn by the Standard Model, presumably not much below the presently explored surface.

In this talk I shall not summarize the conference by going in more detail through the many topics discussed at the conference. Let me just list the main themes. After we were reminded by Wolfenstein that $CP$ violation is now with us for 36 years, we heard in the talks of Roos and Sagawa the news about the $B$–decay measurements from BaBar and Belle. The results for $\sin 2\beta$ have still large errors, but they provide clear signs that $CP$ is violated also in the $B$–system. For the first time $CP$–violation has been observed outside the $K$–system. Further results for $B$–decays came in from CLEO, as reported by Gao. In the Standard Model $CP$ violation arises as a by-product of flavor mixing. But any extension of the model, e. g. towards supersymmetric theories, has its new sources of $CP$–violation, as discussed by Chang.

Flavor physics cannot be seen disjointly from other parts of particle physics, in particular from QCD or from extensions of the Standard Model towards a deeper understanding of gravity (see the talks of Liu and Li on chiral symmetry and of Kim on the mysteries surrounding the cosmological constant).

As we heard in the talks of Koepke and Hsiung, direct $CP$–violation, which is expected in the Standard Model, seems to be established both at CERN and at FNAL, although the violation in the US is larger than in Europe. The chairman of this conference, Yue–Liang Wu, described in detail the present theoretical pricture, which seems to suggest that the "true" value for direct $CP$–violation is about in the middle between the FNAL and CERN results. $CP$–violation has not been observed, thus far, for baryons. This might change in the future, as pointed out by Valencia in his talk on hyperon decays.

Besides $CP$–violation, there are many other features to be studied in the physics of charmed and $b$–flavored particles (see the reviews by Kutschke on the FNAL results, of Lista on BaBar, and of Antilogus on the results from Delphi).

The decays of $B$–mesons provide us with a beautiful testing ground to study the interplay between QCD and flavor dynamics. This interplay was discussed in the talks of Cheng, Chiu and Lu.

Although flavor physics is the corner of the Standard Model which is very close to the experiments, it is not immune with respect to extrapolations of the Standard Model. Thus far the experiments have not provided a direct hint to where the exit road which takes us beyond the Standard Model is leaving, but this could change soon, as discussed by Ali and Masiero with respect to the exit towards supersymmetry and by Ng with respect to the path leading to the jungle of extra dimensions.

New results from BES were discussed by Liu. The present situation at LEP concerning the still hidden "Higgs" particle was outlined by Jin. Yuan described the interplay between the physics of the $t$–quark and of the "Higgs" particle. The future of the $t$–quark physics, from the TEVATRON to the LHC and LC, was discussed by Yeh.

During the second half of the nineties, a new field of flavor physics has opened its doors, the field of lepton flavor mixing, most notably seen in the progress which was made in the study of neutrino oscillations. One should be reminded that neutrino oscillations were first discussed by Pontecorvo about 40 years ago in connection with the $K^0 - \bar{K}^0$–oscillations. Thus they came up in close contact with the flavor physics of quarks. We heared from Kaneyuki about the impressive progress made in Kamioka. Neutrino oscillations seem to be firmly established, although many details, in particular the absolute magnitude of the neutrino masses, are still unknown. Like in the $K^0 - \bar{K}^0$– system we know the mass splittings among the mass eigenstates much better than the masses themselves.

Presumably the only way to find out more about the absolute magnitude of the neutrino masses in the laboratory is to study the double $\beta$ decay.

Unfortunately, we could not hear about the future plans in this field, since professor Klapdor–Kleingrothaus could not come. Of course, double $\beta$–decay is only feasable as a tool to investigate the neutrino mass matrix if neutrinos are either pure Majorana particles, or mixtures of Dirac and Majorana states; in case of massive Dirac states there would be no effect, due to the lepton number conservation.

In the past neutrino physics was a part of high energy physics in which neutrinos were used as tools to study the structure of nuclear matter, like the quark structure functions of the nucleon. Now the focus has changed. The neutrinos themselves are the subject of investigation. It might well be that the new insights obtained in this field allow us to find important information about the dynamics inside the lepton sector. I do not have to stress how important it would be to confirm the results about neutrino mixing obtained in studying the solar and atmospheric neutrinos by laboratory measurements. Longley discussed this in connection to the Minos project in the US, Suzuki in connection to the Kamland projekt in Japan.

Neutrinos are special elementary objects in the sense that they are electrically neutral. Thus the dynamics of the neutrinos can easily be influenced by phenomena beyond the Standard Model, e. g. by mixing with states which have no residence permit within the framework of the Standard Model. Unified Gauge theories, based on the gauge group $SO(10)$, are good examples for this phenomenon.

It is well–known that within a theory based on $SO(10)$[3] one is able to describe a nontrivial mass and mixing pattern for the neutrinos. Moreover, the simplest schemes for the breaking of the symmetry suggest simple relations between the masses of the quarks and the leptons, in particular between the masses of the charged leptons and the down–type quarks[4].

Furthermore, the see–saw mechanism to generate the neutrino masses can easily be implemented in the $SO(10)$–framework. It connects the flavor mixing in the quark sector (typically described by small mixing angles) and in the lepton sector. However, this connection can only be made if something is known about the mass and mixing pattern of the righthanded massive Majorana partners of the observed neutrinos. A large mixing between the light neutrinos is possible, if the structure of the mass matrix for the righthanded Majorana states is similar to the mass–matrices of the quarks and charged leptons[5].

Recently much interest has been devoted to the study of additional large dimensions[6], i. e. dimensions, which go beyond the Minkowski (3+1)–structure. They are possible, provided all fields of the Standard Model propagate only in the four–dimensional subspace, but fields which are allowed with respect to the S.M. gauge group, in particular gravitons and righthanded neutrinos, are singlets to propagate in a larger space–time manifold. Departures from the inverse–square law of Newton are expected in this case, but have not been seen down to the submillimeter scale. Nevertheless, if such extra dimensions are there, the fundamental energy scale associated with gravity will not be the Planck scale of about $10^{19}$ GeV, but could be much lower. In particular for one extra dimension the scale is expected to be about $10^8$ GeV, in which case neutrino physics is likely to be the only possibility to find out something about the extra dimension, as discussed by Lam. A righthanded neutrino would be derived from a 5–dimensional Dirac field, and it would act like a sterile neutrino or a tower of sterile neutrinos mixed with the ordinary neutrinos. The oscillation pattern of the observed neutrinos can be quite different to the one of the extended Standard Model with only three massive neutrinos. One feature of those models is that a considerable amount of the active neutrino flux is dispersed into sterile neutrinos. In the case of $\nu_e$ and $\nu_\mu$ neutrinos there are good limits for such an effect, and no dispersion is seen thus far. However, no limit exists for the $\tau$–neutrinos. In the foreseeable future the experimentalists will provide us with enough data to set rather stringent limits on a possible diversion of neutrino flux into the dark corner of extra dimensions, and this will at the same time provide strong constraints on the physics of extra dimensions. Less likely, but certainly possible, is their actual discovery in "looking through the neutrino glas" beyond the realm of our four–dimensional world.

The phenomenon of flavor mixing is an intrinsic part of the Standard Model, but the part whose dynamics is not understood. The world would be simpler without flavor mixing, but nature seems to prefer to go off the simplest road. Obviously the mixing between the families is intrinsically related to the dynamics of quark mass generation. The observed fact that the flavor mixing angles in the quark sector are small must be related to the strong mass hierarchy observed in the mass spectrum.

The "standard" parametrization of the flavor mixing matrix (advocated by the Particle Data Group) and the original Kobayashi–Maskawa parametrization[7] were introduced without taking possible links between the quark masses and the flavor mixing parameters into account. The parametrization Xing and I introduced some time ago (for a review see ref. (5)) is based on such a connection, although the specific relations between flavor mixing angles and quark masses might be more complicated than commonly envisaged. It is a parametrization which allows to interpret the phenomenon of flavor mixing as an evolutionary or tumbling process. In the limit in which the masses of the light quarks $(u, d)$ and the medium light quarks $(c, s)$ are set to zero, while the heavy quarks $(t, b)$ acquire their masses, there is no flavor mixing. Once the masses of the $(c, s)$–quarks are introduced, while the $(u, d)$–quarks remain massless, the flavor mixing is reduced to an admixture between two families, described by one angle $\Theta$. As soon as the $u$– and $d$–quark masses are introduced as small perturbations, the full flavor mixing matrix involving a complex phase parameter and two more mixing angles $(\Theta_u, \Theta_d)$ appears. These angles can be interpreted as rotations between the states $(u, c)$ and $(d, s)$, respectively. In either the "standard" parametrization or the Kobayashi–Maskawa representation, however, such specific limits are difficult to consider. The representation I prefer is given by:

$$
\begin{aligned}
V &= \begin{pmatrix} c_u & s_u & 0 \\ -s_u & c_u & 0 \\ 0 & 0 & 1 \end{pmatrix} \begin{pmatrix} e^{-i\varphi} & 0 & 0 \\ 0 & c & s \\ 0 & -s & c \end{pmatrix} \begin{pmatrix} c_d & -s_d & 0 \\ s_d & c_d & 0 \\ 0 & 0 & 1 \end{pmatrix} \\
&= \begin{pmatrix} s_u s_d c + c_u c_d e^{-i\varphi} & s_u c_d c - c_u s_d e^{-i\varphi} & s_u s \\ c_u s_d c - s_u c_d e^{-i\varphi} & c_u c_d c + s_u s_d e^{-i\varphi} & c_u s \\ -s_d s & -c_d s & c \end{pmatrix},
\end{aligned}
\tag{4}
$$

where $s_u \equiv \sin \Theta_u$, $c_u \equiv \cos \Theta_u$, etc. The three mixing angles can all be arranged to lie in the first quadrant, i. e., all $s_u, s_d, s$ and $c_u, c_d, c$ are positive. The phase $\varphi$ may in general take all values between 0 and $2\pi$. Clearly $CP$ violation is present, if $\varphi \neq 0$ or $\varphi \neq \pi$.

In many models for the quark mass matrices there exist simple relations between the mass eigenvalues and the mixing angles $\Theta_u$ and $\Theta_d$:

$$
\tan \Theta_u = |V_{ub}/V_{cb}| \approx \sqrt{m_u/m_c}
$$

$$
\tan \Theta_d = |V_{td}/V_{ts}| \approx \sqrt{m_d/m_s}
\tag{5}
$$

The typical estimates of the quark masses give $\sqrt{m_u/m_c} \sim 0.06 \ldots 0.08$, a value which is slightly lower than the observed ratio $|V_{ub}/V_{cb}| \approx 0.09 \pm 0.02$.

The angle $\Theta_d$ is determined rather precisely by the ratio $\sqrt{m_d/m_s}$, if one takes the results of chiral symmetry breaking into account. One expects $|V_{td}/V_{ts}| \cong 0.22 \ldots 0.23$.

In the representation I am advocating the mixing strength between the first and second generation is determined by the two mixing angles $\Theta_u$ and $\Theta_d$. Both angles vanish, if the masses of the light quarks $m_u$ and $m_d$ are turned off. In this limit $CP$–violation would not be present. Suppose only one of the light quarks $u \sim d$ acquires a mass. In this case both $\Theta_u$ and $\Theta_d$ are nonzero, but one of the angles is extremely small. Its magnitude depends on the actual structure of the mass matrix. In specific models[5] $\Theta_u$ is of the order of $\sqrt{m_d/m_s} \cdot (m_s/m_b)$, if $m_u$ is zero, i. e. about $10^{-2}$, an order of magnitude smaller than observed. Likewise $\Theta_d$ is of the order of $\sqrt{m_u/m_c} \cdot (m_c/m_t)$, if $m_d$ is zero, i. e. about two orders of magnitude smaller than the observed value. Thus the observed value of the flavor mixing angles $\Theta_u, \Theta_d$ give a strong hint that neither $m_u$ nor $m_d$ are vanishing.

The mixing element $V_{us}$ is given by:

$$V_{us} \cong s_u - s_d e^{-i\varphi} \tag{6}$$

where $s_{u,d} = sin\Theta_{u,d}$. Thus a precise determination of $V_{us}$, along with a precise determination of $s_u$ and $s_d$, would allow to determine the phase $\varphi$ responsible for $CP$–violation.

In the mass matrix models mentioned above $\Theta_u$ and $\Theta_d$ are given by $\sqrt{m_u/m_c}$ and $\sqrt{m_d/m_s}$ respectively. The relation for $V_{us}$ fixes a triangle in the complex phase, which is congruent to the unitarity triangle[5].

It is well–known that the absolute value of the Cabibbo transition $V_{us}$ is essentially identical to $\sqrt{m_d/m_s}$, which can be determined very well from the chiral dynamics of QCD. Thus there is little space for the contribution from $s_u$, and one concludes that the $CP$–violating phase $\varphi$ must be close to $90^0$, a situation which can be described as maximal $CP$–violation[5].

The picture which emerges is the following: In the absence of the $u$– and $d$– masses only one mixing angle $\Theta$, describing the mixing between the second and third family, is present. At the second step the mixing angle $\Theta_d \approx \sqrt{m_d/m_s}$ appears, while $\Theta_u \approx \sqrt{m_u/m_s}$ can be introduced in a third step. The Cabibbo transition $V_{cd}$ is then given by

$$V_{cd} \approx \Theta_d - \Theta_u e^{-i\varphi} \approx \sqrt{\frac{m_d}{m_s}} - \sqrt{\frac{m_u}{m_c}} e^{-i\varphi} \tag{7}$$

$CP$ violation is clearly seen as a phenomenon related to the generation of mass for the first family. As mentioned above, the observed absolute magnitudes of $V_{us}$ and $V_{cd}$ agree with the ratio $\sqrt{m_d/m_s}$, and the correction coming from $\sqrt{m_u/m_c}$ cannot be sizeable, implying that the phase angle $\varphi$ must be close to $90^0$. In this case one can rewrite $V_{cd}$ as follows:

$$V_{cd} \approx \sqrt{\frac{m_d}{m_s}} + \sqrt{\frac{-m_u}{m_c}} \tag{8}$$

i. e. the mass of the $u$–quark enters with a negative sign, such that a phase angle of $90^0$ appears. This phase angle of $90^0$ might be a signal for a specific discrete symmetry. Since the phase angle $\varphi$ corresponds to the angle $\alpha$ in the unitary triangle, it is implied that the unitarity triangle is rectangular. The other two angles $\beta$ and $\gamma$ are given in terms of the two mass ratios $\sqrt{m_u/m_c}$ and $\sqrt{m_d/m_s}$. Taking the central values of the quark masses, one finds $\beta \approx 20°(sin2\beta \approx 0.64)$ and $\gamma \approx 70^0$. More specifically, the range of $\sin 2\beta$ varies between 0.56 and 0.70, as we vary $\sqrt{m_u/m_c}$ in the most likely range 0.06 ... 0.08. The experimental data from BaBar and Belle are consistend with this range of values for $\sin 2\beta$, but differ so much from each other that a clear conclusion cannot be drawn. Nevertheless the value $\sin 2\beta \approx 0.70$ seems to me the upper edge of the allowed range. If the experiments eventually give a higher value, the theoretical basis of connecting the flavor mixing angles with quark mass ratios, as discussed becomes questionable.

But even if this would be the case, which I doubt, we must conclude that the success of the Standard Model with respect to $CP$ violation is impressive. Its strength both is the $K$ meson sector and in the $B$ meson sector is predicted by parameters which on their own have nothing to do with $CP$ violation, but rather with flavor violation, the flavor mixing angles. The phase parameter $\varphi$ describing the $CP$–violation is large; it might even be $90^0$. In view of this success it seems unlikely to me that the observed $CP$ violation comes from something else than the mechanism offered for free by the Standard Model, where it is linked to the flavor mixing. However, small deviations from the Standard Model expectations might be there and should be searched for in the future. Whether they are there and whether they are specific enough to point towards a specific model, like the supersymmetric extension of the Standard Model, remains to be seen.

Closing this conference, let me thank, especially also in the name of all foreign participants, the organizing committee and in particular its chairman, Prof. Yue–Liang Wu, for taking on the complicated task to organize this conference in this wonderful and remote place.